\documentclass[12pt,preprint]{aastex}

\usepackage{rotating}

\usepackage{booktabs,threeparttable}

\slugcomment{To appear in Apj}

\shorttitle{BLR Micro and millilensing}

\shortauthors{Guerras et al.}

\begin{document}

\title{Microlensing of Quasar UV Iron Emission}

\author{E. GUERRAS\altaffilmark{1,2}, E. MEDIAVILLA\altaffilmark{1,2}, J. JIMENEZ-VICENTE\altaffilmark{3,4}, C.S. KOCHANEK\altaffilmark{5}, J. A. MU\~NOZ\altaffilmark{6}, E. FALCO\altaffilmark{7}, V. MOTTA\altaffilmark{8}, K. ROJAS\altaffilmark{8}}

\altaffiltext{1}{Instituto de Astrof\'{\i}sica de Canarias, V\'{\i}a L\'actea S/N, La Laguna 38200, Tenerife, Spain}

\altaffiltext{2}{Departamento de Astrof\'{\i}sica, Universidad de la Laguna, La Laguna 38200, Tenerife, Spain}

\altaffiltext{3}{Departamento de F\'{\i}sica Te\'orica y del Cosmos, Universidad de Granada, Campus de Fuentenueva, 18071 Granada, Spain}

\altaffiltext{4}{Instituto Carlos I de F\'{\i}sica Te\'orica y Computacional, Universidad de Granada, 18071 Granada, Spain}

\altaffiltext{5}{Department of Astronomy and the Center for Cosmology and Astroparticle Physics, The Ohio State University, 4055 McPherson Lab, 140 West 18th Avenue, Columbus, OH, 43221 }

\altaffiltext{6}{Departamento de Astronom\'{\i}a y Astrof\'{\i}sica, Universidad de Valencia, 46100 Burjassot, Valencia, Spain.}

\altaffiltext{7}{Center for Astrophysics, 60 Garden Street, Cambridge, MA 02138, USA}

\altaffiltext{8}{Departamento de F\'{\i}sica y Astronom\'{\i}a, Universidad de Valpara\'{\i}so, Avda. Gran Breta\~na 1111, Valpara\'{\i}so, Chile}

%---------------------------------------------------------------------

% 0. ABSTRACT

\begin{abstract}

We measure the differential microlensing of the UV Fe II and Fe III emission line blends between 14 quasar image pairs in 13 gravitational lenses. We find that the UV iron emission is strongly microlensed in 4 cases with amplitudes comparable to that of the continuum. Statistically modeling the magnifications we infer a typical size of $r_s\sim 4\sqrt{M/M_\odot}$ light-days for the Fe line emitting regions which is comparable to the size of the region generating the UV continuum ($\sim3-7$ light-days). This may indicate that a significant part of the UV Fe II and Fe III emission originates in the quasar accretion disk.

\end{abstract}


%---------------------------------------------------------------------

%---------------------------------------------------------------------

\section{Introduction}

Iron, the stable end product of nucleosynthesis, has a large number of energy levels, generating thousands of emission line transitions distributed throughout the UV and optical bands that likely makes Fe II the main emission line contributor to the overall spectra of quasars and AGN.  Statistical studies of quasar spectra find that variations between them are dominated by the relative strength of the Fe II emission (Boroson \& Green 1992). In spite of the relevance of the iron lines to understanding the physics of AGN, both the mechanism generating the Fe II emission (Ferland et al. 2009), and the spatial scale of the region where it is emitted (Barth et al. 2013) are poorly understood. Two bands of Fe II emission are usually studied (see e.g. Baldwin et al. 2004): the UV  pseudo-continuum between C III]$\lambda1909$ and Mg II$\lambda2798$ and the optical blends in the H$\gamma$-H$\beta$ region. The few reverberation mapping studies of Fe II indicate that the region emitting the UV Fe II lines (Maoz et al. 
1993) is considerably smaller than the region emitting the optical Fe II lines (Kuehn 2008, Barth et al. 2013).

The size of the region generating the broad emission lines (BEL) in quasars can be also inferred from the impact of microlensing on the BEL. In a multiply imaged quasar, the magnification of each image of the quasar varies with time due to lensing by the stars in the lens galaxy (see the review by Wambsganss 2006). The dependence of the microlensing magnification on the size of the emission region has been used to estimate the size of different quasar regions including the accretion disk (see, e.g. Pooley et al. 2007, Morgan et al. 2010, Blackburne et al. 2011, Jim\'enez-Vicente et al. 2012), the Broad Line Region (BLR) (Abajas et al. 2007, Sluse et al. 2012, Motta et al. 2012, Guerras et al. 2013) and the non-thermal X-Ray emission region (Pooley et al. 2007; Morgan et al. 2008, 2012; Chartas et al. 2009; Dai et al. 2010; Blackburne et al. 2011; Mosquera et al. 2013).  In the particular case of the iron emission lines, Sluse et al. (2007) analyzed spectra of RXS J1131$-$1231 and found that a large fraction 
of the optical Fe II emission arises in the outer parts of the BLR,  although they also found evidence of a very compact region associated with Fe II. Evidence of significant microlensing of the UV Fe II emission was also found in Q2237+0305 (Sluse et al. 2011).

Here we study the microlensing of the UV iron emission in a sample of 14 pairs of lensed quasar images, combining the spectra compiled by Mediavilla et al. (2009) with new observations. In Section 2, we describe the data and the procedure used to isolate the Fe II and Fe III line emission from the continuum and to then measure its microlensing. In Section 3, we use the measured microlensing amplitudes to derive constraints on the size of the UV iron emitting region and we discuss and summarize the results in Section 4.

\section{Data Analysis \label{DA}}

We started with the sample compiled by Mediavilla et al. (2009) and then added unpublished archival spectra taken with the VLT or the MMT, as summarized in Table \ref{tbl-1}. We focus on the wavelength region between the C III]$\lambda 1909$ and Mg II$\lambda 2798$ emission lines and we will use these lines to define a flux ratio baseline that is only weakly affected by microlensing (Mediavilla et al. 2009, 2011a, Guerras et al. 2013). Table \ref{tbl-2} defines the wavelength regions we consider. We loosely follow the definition of the Fe II wavelength windows by Francis et al. (1991). The primary modification is that we do not include the regions around C III]$\lambda 1909$ to avoid modeling the blended emission. Iron emission is split into two windows designated Fe(1), dominated by Fe III, and Fe(2),  dominated by Fe II (Vestergaard \& Wilkes 2001), each bracketed by line-free continuum regions (Kuraszkiewicz et al. 2002, Francis et al. 1991, Brotherton et al. 2001). Fe(1) corresponds to 2050-2115\AA\, and 
Fe(2) corresponds to three regions: 2250-2320\AA, 2333-2445\AA, and 2470-2625\AA. 

We first fit 4 straight lines to the continuum regions bracketing the Fe emission windows (plus a continuum region bluewards of the C III]$\lambda 1909$ line and other redwards of the Mg II$\lambda 2798$ line) and subtract it from the spectrum. Then, for each image-pair we normalize the continuum subtracted spectra to match the core of the Mg II$\lambda 2798$ emission lines defined by the total flux within $\pm$ FWHM/2 of the line center. Provided these low ionization lines are only weakly affected by microlensing, as we found in Guerras et al. (2013), the normalization constant (that is, the ratio between the Mg II$\lambda 2798$ emission lines) gives us the intrinsic macrolens magnification between the images. The flux ratio of the continuum as compared to that of the Mg II$\lambda 2798$ emission lines then gives us an estimate of the continuum microlensing magnification:

\begin{equation}
\Delta m_{\rm cont} = (m_1-m_2)_{\rm cont} - (m_1-m_2)_{\rm MgII\lambda 2798}.
\end{equation}
In Figure \ref{colorines}, the superposition of the continuum subtracted and normalized (to the Mg II$\lambda 2798$ emission line) spectra are shown for each image-pair. The average SDSS quasar  spectrum (Vanden Berk et. al. 2001), continuum-subtracted following the same procedures, is shown for comparison. In all cases (except SBS 0909+532 and SDSS J1353+1138) the Mg II$\lambda 2798$ based normalization also matched the C III]$\lambda 1909$ emission lines well, which shows that the continuum subtraction procedure has worked well and that there are no significant effects due to differential atmospheric refraction or slit misalignments. The exceptions were  SBS 0909+532 which is strongly affected by differential extinction in the lens galaxy (Motta et al. 2002, Mediavilla et al. 2005, Mediavilla et al. 2011b) and, at a lower level, SDSS J1353+1138. Assuming that the differences in normalization between the C III]$\lambda 1909$ and the Mg II$\lambda 2798$ emission lines in these two objects arise from 
extinction we have applied a linear extinction correction to match both emission lines simultaneously. 

Examining Figure \ref{colorines}, we see significant differences between the spectra in the region between C III]$\lambda 1909$ and Mg II$\lambda 2798$ in 4 cases: SDSS J0806+2006AB, FBQS J0951+2635AB, QSO 0957+561AB and SDSS J1353+1138. For each pair of images, 1 and 2, we can estimate the microlensing in the two Fe regions by comparing the differential flux ratios between the iron blends and the Mg II$\lambda2798$ line that sets the baseline for no microlensing magnification. For example, we define

\begin{equation}
\Delta m_{\rm Fe(1)} = (m_1-m_2)_{\rm Fe(1)} - (m_1-m_2)_{\rm MgII\lambda 2798},
\end{equation}
for region Fe(1) and similarly $\Delta m_{\rm Fe(2)}$ for region Fe(2). The same analysis can be performed using C III]$\lambda1909$ as the unmicrolensed baseline. For C III]$\lambda 1909$ the definition of the continuum below the line is less well defined than for Mg II$\lambda2798$, so these results should be treated with more caution. Using the C III]$\lambda 1909$ line we can, then, define estimates of the microlensing amplitudes $\Delta m'_{\rm cont}$, $\Delta m'_{\rm Fe(1)}$, and $\Delta m'_{\rm Fe(2)}$. All the resulting microlensing estimates are presented in Tables \ref{tbl-4} and \ref{tbl-5}.

\begin{figure}[h]

%\plotone{fig2flux_sin_v5.pdf}
\includegraphics[width=1.0\textwidth]{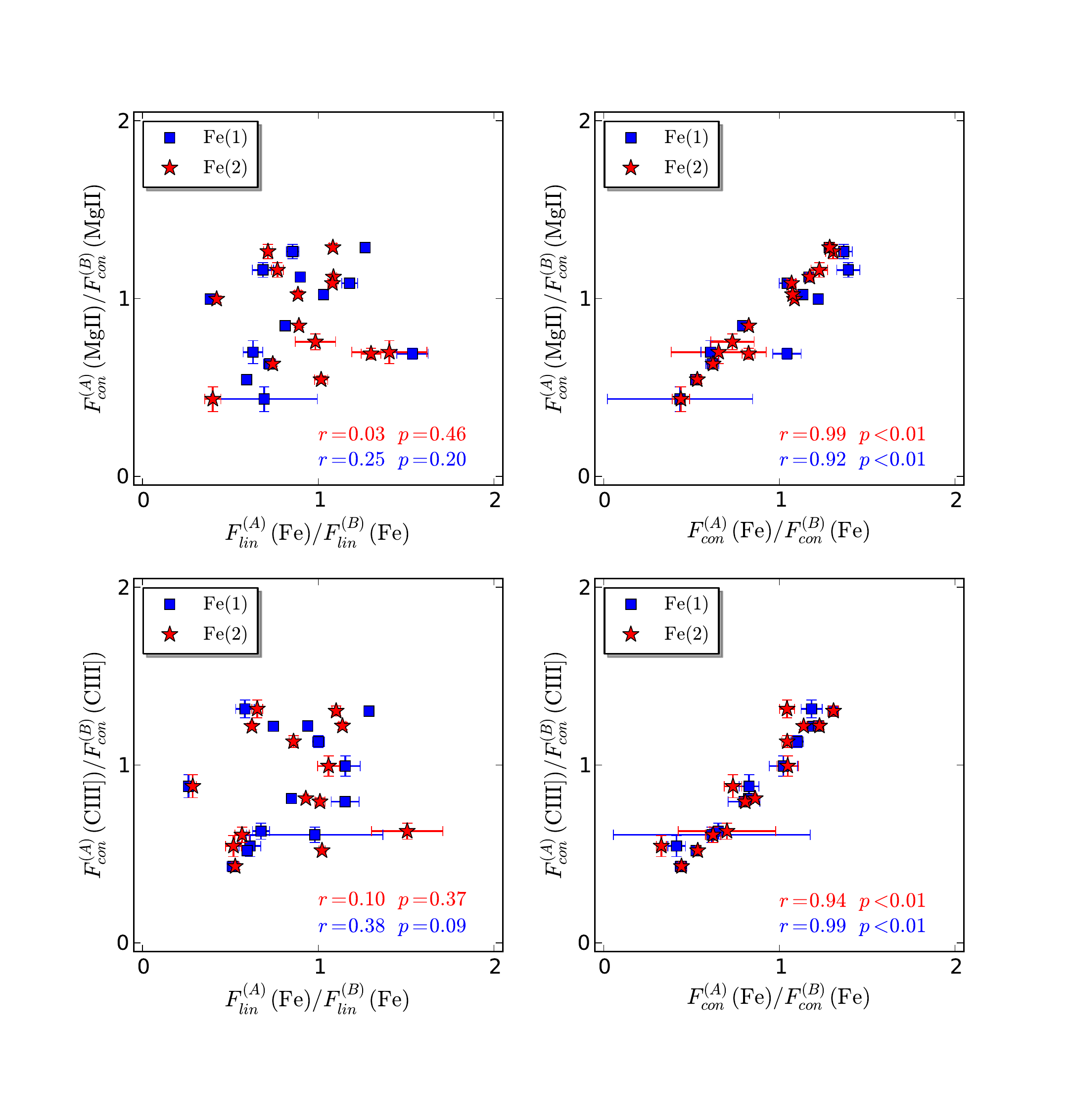}

\caption{Comparison of Fe(1) and Fe(2) emission line and continuum flux ratios with the Mg II$\lambda2798$ (top) and C III]$\lambda 1909$ continuum ratios (bottom). Blue squares correspond to Fe(1) and red stars correspond to Fe(2). \label{scatter}}

\end{figure}

In Figure \ref{scatter} we compare the microlensing magnification estimates for the continuum underlying each line or blend, C III]$\lambda 1909$, Mg II$\lambda2798$, Fe(1) and Fe(2), finding very good linear covariances. The Pearson correlation coefficients are above $0.92$ in all cases, with one-tailed p-values well under $0.01$. In the same Figures we also compare the microlensing measured in the Fe(1) and Fe(2) line regions with the microlensing of the continuum regions adjacent to the C III]$\lambda 1909$ and Mg II$\lambda2798$ emission lines. We find that the Fe(1) line region has a low degree of correlation with the continuum, while the Fe(2) line region is uncorrelated.

\section{Constraining the Size of the UV Iron Emission Line Region}

We follow the procedure we used in Guerras et al. (2013)  to estimate the size of the emission regions corresponding to Fe(1), Fe(2) and the continuum regions adjacent to the Mg II$\lambda2798$ and C III]$\lambda 1909$ emission lines. We start by computing microlensing magnification maps using the inverse polygon mapping technique (Mediavilla et al. 2006, Mediavilla et al. 2011a). We take the dimensionless surface density, $\kappa$, and shear, $\gamma$, for each image from the lens models by Mediavilla et al. (2009) and Mediavilla et al. (2011). We assume a mass fraction in stars of 5\% (Abajas et al. 2007, Mediavilla et al. 2006, Pooley et al. 2009, 2012) and a stellar mass of $M=1M_\odot$. We generate microlensing magnification maps with an outer scale of 1100 light-days and with a pixel scale of 0.04 Einstein radii (equal to 0.6 light-days in the worst case). Each magnification map is unit normalized and convolved with a Gaussian of size, $r_s$,  $I\propto \exp(-R^2/2r^2_s)$ to model the source. We 
consider a 
linear (logarithmic) grid of sizes from $r_s=1.5$ to $13$ light-days with steps $\Delta r_s=0.5$ light-days ($\Delta \log_{10}{r_s}=0.0408$). The probability of observing a magnitude difference $\Delta m_{obs,k}$ for image pair $k\ (k=1,...,14)$ given a source size $r_s$ is then

\begin{equation}
p_k(\Delta m_{obs,k}|r_s)\propto \int{f_{r_s,k,1}(m_1)f_{r_s,k,2}(m_1-\Delta m_{obs,k})dm_1}
\end{equation}
where $f_{r_s,k,1}(m)$ and $f_{r_s,k,2}(m)$ are the frequency histograms obtained from the convolved magnification maps for images 1 and 2 (of pair $k$), respectively. The joint likelihood for all the image pairs,

\begin{equation}
L(\Delta m_{obs,1},...,\Delta m_{obs,14}|r_s)=\prod_{k=1}^{14}{p_k(\Delta m_{obs,k}|r_s)},
\end{equation}
then gives the likelihood distribution for the size $r_s$. 

%\begin{equation}
%p(r_s|\Delta m_{obs,1},...,\Delta m_{obs,14})= {L(\Delta m_{obs,1},...,\Delta m_{obs,14}|r_s)\over %\sum_{r_s} L(\Delta m_{obs,1},...,\Delta m_{obs,14}|r_s)}.
%\end{equation}

\begin{figure}[h]

%\plotone{pink_linearprior_linearaxis_mle_v5.pdf}
\includegraphics[width=1.0\textwidth]{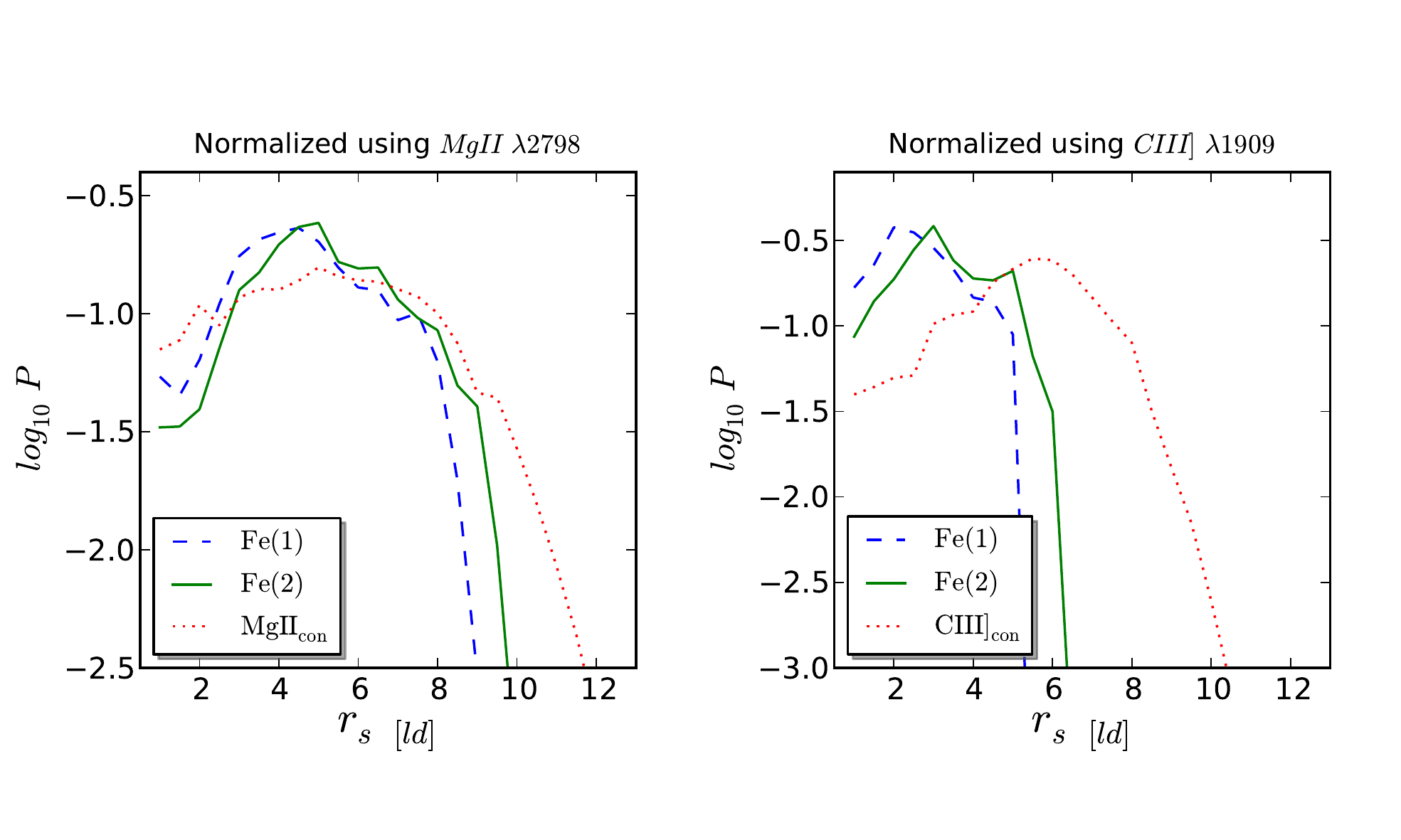}

\caption{Posterior probabilities for a uniform prior on $r_s$. In the left (right) panel, the baseline for no microlensing magnification was set using the Mg II$\lambda 2798$ (C III]$\lambda 1909$) emission line. The dashed, solid and dotted curves correspond to Fe(1), Fe(2) and the continuum region associated with the normalizing line, Mg II$\lambda 2798$ (left) or C III]$\lambda 1909$ (right). \label{mle}}

\end{figure}

\begin{figure}[h]

%\plotone{pink_log10prior_log10axis_mle_v5.pdf}
\includegraphics[width=1.0\textwidth]{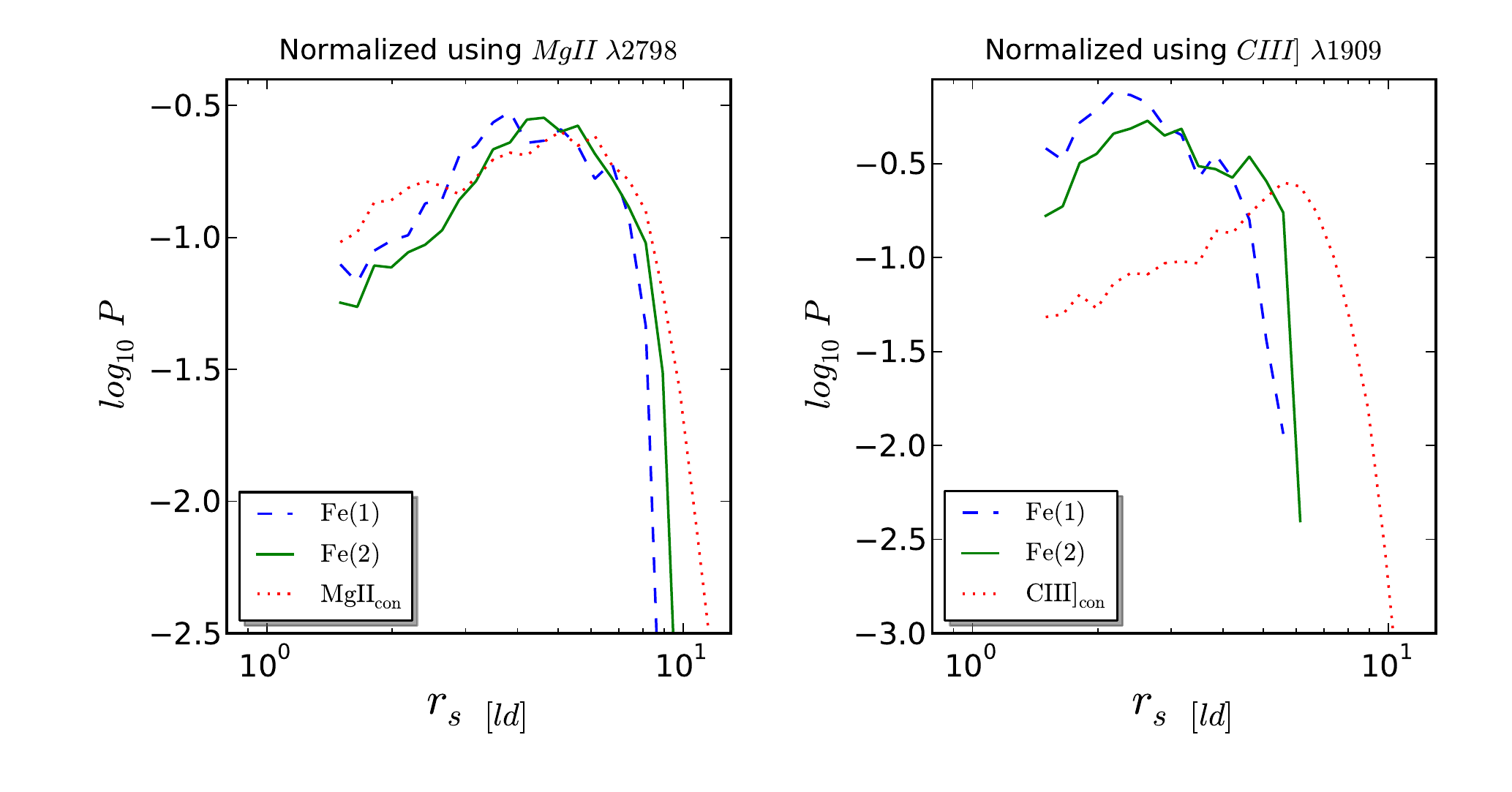}

\caption{Same as Figure \ref{mle} but with a logarithmic prior on $r_s$. \label{logmle}}

\end{figure}

Normalizing the likelihood functions to unity gives the Bayesian posterior probabilities with either a uniform (linear grid) or logarithmic (log grid) prior on $r_s$. Figures \ref{mle} and \ref{logmle} show, for linear and logarithmic grids respectively, the resulting probability distributions for the Fe(1) and Fe(2) line regions, and the continuum under the Mg II$\lambda 2798$ (C III]$\lambda1909$) line when using the Mg II$\lambda 2798$ (C III]$\lambda1909$) line to estimate the flux ratios in the absence of microlensing. The most significant result is that the UV iron blends seem to originate in a region of size comparable to that of the underlying UV continuum. From these posterior probability distributions we obtain size estimates, in $\sqrt{M/M_{\odot}}$ light-day units, for the uniform (logarithmic) prior of $r_s=5.3\pm 2.4$ ($r_s=5.3\pm 2.1$) and $r_s=5.3\pm 1.8$ ($r_s=5.3\pm 1.7$) for the Mg II$\lambda2798$ and C III]$\lambda 1909$ continua, respectively, in reasonable agreement with current 
expectations about the size 
of the region generating the continuum in quasars (e.g. Morgan et al. 2010, Jim\'enez-Vicente et al. 2012). We obtain similar sizes, in $\sqrt{M/M_{\odot}}$ light-day units, for the Fe(1) and Fe(2) line emission regions with $r_s=4.6\pm 1.8$ ($r_s=4.8\pm 1.7$) and $r_s=5.1\pm 1.8$ ($r_s=5.1\pm 1.7$), using the Mg II$\lambda2798$ lines as the magnification reference and $r_s=2.7\pm 1.1$ ($r_s=2.9\pm 0.9$) and $r_s=3.3\pm 1.2$ ($r_s=3.4\pm 1.1$), using C III]$\lambda 1909$. While the C III]$\lambda 1909$ estimates are systematically smaller, the results are statistically consistent.

\section{Discussion and Conclusions}

%The main result inferred directly from the spectra, is the presence in the case of 4 quasar image-pairs of strong differences in the shape of the profiles of the UV iron blends that we have interpreted in terms of differential microlensing.

We have found evidence that the UV iron line pseudo-continuum regions are microlensed, with an amplitude comparable to that of nearby continuum emission regions. When we formally estimate the size we find $r_s\sim 4\sqrt{M/M_\odot}$ light-days, slightly smaller than the continuum regions ($r_s\sim 3-7\sqrt{M/M_\odot}$ light-days) and far smaller than either the high or low ionization line emission regions in the BLR as estimated either with microlensing (Guerras et al. 2013) or reverberation mapping (see e.g. Zu et al. 2011). These quantitative results should be regarded as preliminary since the sample is small and a single object (SDSS J1353+1138) has a disproportionate impact on the size estimates. In any case, our estimate for the size  is in reasonable agreement with the results for the UV Fe II emission region based in reverberation mapping by Maoz et al. (1993). Other reverberation mapping studies indicate that the Fe II optical emission lines arise from a substantially larger region (Kuehn 2008, Barth 
et al. 
2013).  However, Sluse et al. (2007) also found that a part of the optical Fe II emission may originate in a more compact region.

%The main result inferred from the comparison between spectra of quasar image-pairs, is the strong difference in the shape of the iron blends found in 4 objects of the sample that we have interpreted in terms of differential microlensing.  Using the strength of microlensing to estimate sizes, we have found that the UV iron emission originates in a region of size comparable to that of the underlying continuum. In principle it could seem that the UV iron emission may arise from a region even smaller. However, the likelihood functions have been obtained from a relatively small sample of objects and all the results should be regarded with caution.  In particular notice that one object, SDSS J1353+1138, has an important weight in the likelihood functions for Fe(1) and Fe(2).

%We have obtained a typical size of $r_s\sim 4\sqrt{M/M_\odot}$ light-days that is substantially smaller than the size estimated for other broad emission lines using microlensing (24 [55] light-days for the low [high] ionization emission lines according to Guerras et al. 2013) or reverberation mapping (see e.g. Zu et al. 2011). 

\begin{figure}[h]
%\plotone{hybrid_superp_v5.pdf}
\includegraphics[width=1.0\textwidth]{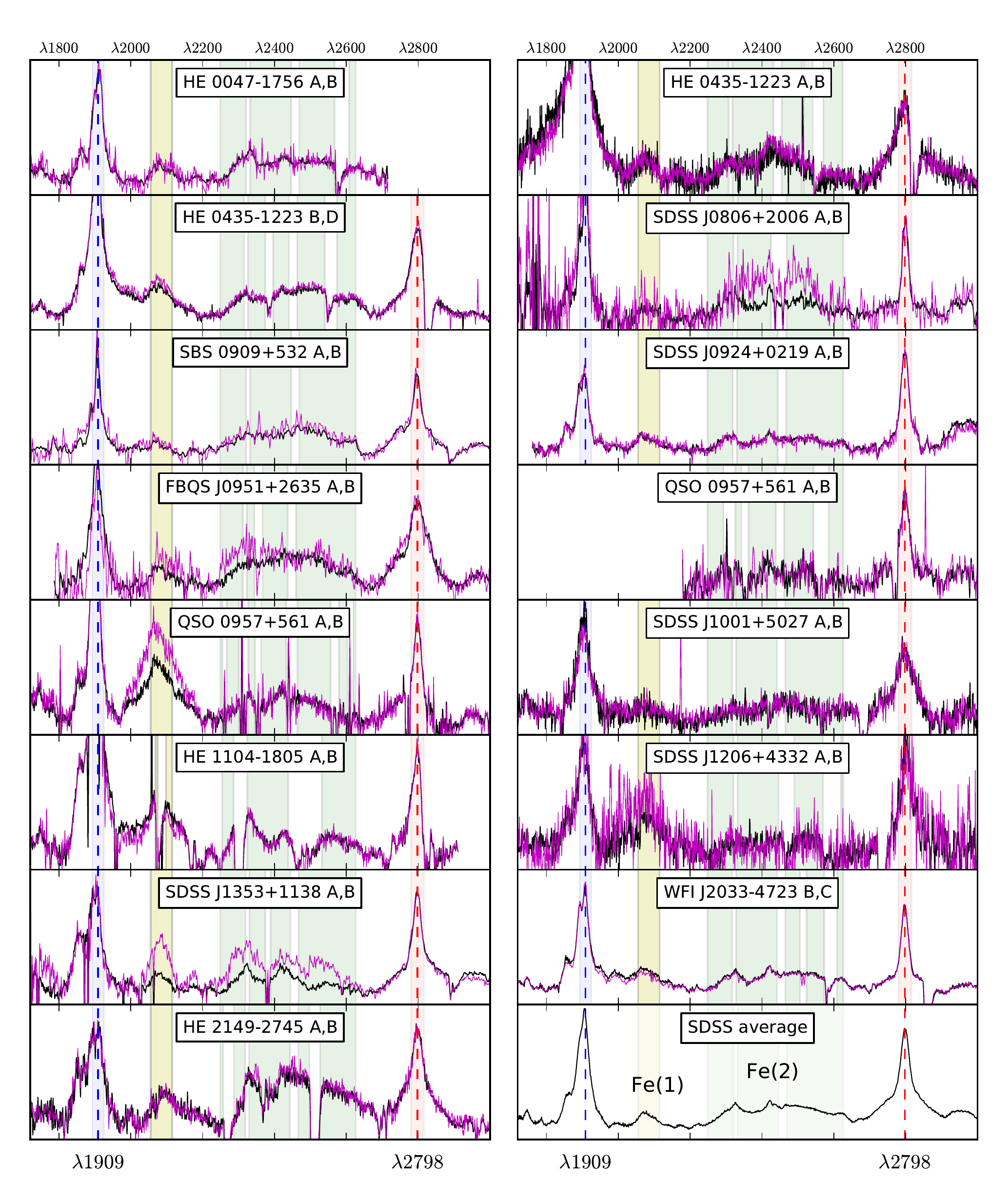}
\caption{
Panels showing superpositions of the paired spectra after continuum subtraction. The shaded regions show
 the wavelength intervals used for the C III] $\lambda 1909$ line core, the Fe(1) blend, the Fe(2)
 blend and the Mg II $\lambda 2798$ line core respectively (see Table \ref{tbl-2}). In the cases of SBS 0909+532 and SDSS J1353+1138
 a linear model was used to correct for differential extinction between the images. \label{colorines}
}
\end{figure}

It is also interesting to explore the shape of the spectra to know whether microlensing acts selectively over some components of the pseudo-continuum and may shed light on the structure and kinematics of the inner regions of quasars. The shape of the unmicrolensed spectra resembles the average SDSS quasar spectra well (see Figure \ref{colorines}). The microlensed spectra, however, seem to be selectively enhanced at some of the spectral features in the iron emission templates from Vestergaard \& Wilkes (2001).  This is particularly true for the Fe(1) blend that appears strongly magnified in 3 of the 4 microlensed objects. On the other hand, in the Fe(2) blend of SDSS J1353+1138 (see also the low S/N spectra from FBQS J0951+2635 and SDSS J0806+2006), the enhanced features look broader and more flat-topped. Notice, in particular, the relative weakness of the C II]$\lambda2326$ emission line compared to the unmicrolensed spectra and the strong enhancement of the Fe II emission at $\sim$2300\AA\ that can be 
hardly identified in the average SDSS quasar spectrum.  A similar relative enhancement of the Fe II emission around the (tentatively identified) Fe III$\lambda2418$ (narrow)+[Ne IV]$\lambda2423$ blend is observed in SDSS J0806+2006. High S/N ratio spectra combined with detailed modeling of the iron emission could help to understand both the origin of the iron emission and the structure of the innermost regions of quasars (inner BLR or/and accretion disk).

%High S/N ratio spectra combined with detailed modeling of the iron emission could help to understand both the origin of the iron emission and the structure of the innermost regions of quasars (inner BLR or/and accretion disk). Evidences for high velocities in the enhanced emission lines (like the needing of broadened templates to model the iron emission) would confirm that part of the iron emission arises from the accretion disk as the small estimate for the size indicates. On the other hand, microlensing variability that can help to scan different regions of the quasar, unresolved with present or even future planned telescopes, can also be present in timescales of a few years. Spectro-photometric monitoring or simply a new observation will be very useful to search for changes since the archival spectra we have used here were taken.

In a series of papers we have used archival spectra of lensed quasars and microlensing to measure the fraction of matter in compact objects (Mediavilla et al. 2009), the size of quasars accretion disk (Mediavilla et al. 2011a, Jim\'enez-Vicente et al. 2012), the size of the BLR (Guerras et al. 2013) and the temperature profile of the quasars accretion disk (Jim\'enez-Vicente et al. 2013). It is clear that the next step is to revisit these objects to search for spectral changes, or even to begin systematic spectroscopic monitoring.

\acknowledgements

This research was supported by the Spanish Ministerio de Educaci\'on y Ciencia with the grants AYA2010-21741-C03-01/02, AYA2011-24728. J.J.V. is also supported by the Junta de Andaluc\'\i a through the FQM-108 project. J.A.M. is also supported by the Generalitat Valenciana with the grant PROMETEO/2009/64. C.S.K. is supported by NSF grant AST-1009756. V.M. gratefully acknowledges support from FONDECYT through grant 1120741.

\clearpage

\begin{deluxetable}{l|llll}

\tabletypesize{\scriptsize}

%\rotate

\centering

\tablecaption{Summary of Data\label{tbl-1}}

%\tablewidth{0pt}

\startdata

\tableline\

Object (pair)     & z & Observation Date  &   Rest wavelenght (\AA)       & Reference \\

\tableline

HE 0047$-$1756  A, B  & 1.67  & 2002 Sep 04 	& (1461  - 2547) & Wisotzki et al., 2004  \\

HE 0435$-$1223  A, B  & 1.689 & 2008 Jan 12  	& (1210  - 3030) & Motta, V., unpublished data  \\

HE 0435$-$1223  B, D  & 1.689 & 2004 Oct 12  	& (1638  - 2996) & Motta, V., unpublished data  \\

SDSS 0806+2006  A, B  & 1.54  & 2005 Apr 12 	& (1575  - 3504) & Inada et al., 2006 \\

SBS 0909+532  A, B  & 1.38  & 2003 Mar 07  	& (0750  - 5695) & Mediavilla et al., 2005  \\

SDSS J0924+0219   A, B  & 1.524 & 2005 Jan 14 	& (1783  - 3170) & Eigenbrod et al., 2006 \\

FBQ 0951+2635   A, B  & 1.24  & 1997 Feb 14 	& (1786  - 4018) & Schecter et al., 1998  \\

QSO 0957+561  A & 1.41  & 1999 Apr 15 		& (0913  - 4149) & Goicoechea et al., 2005  \\

QSO 0957+561  B & 1.41  & 2000 Jun 2-3  	& (0913  - 4149) & Goicoechea et al., 2005  \\

QSO 0957+561  A, B & 1.41  & 2008 Jan 13  	& (1330  - 3380) & Motta et al., 2012  \\

SDSS J1001+5027   A, B  & 1.838 & 2003 Nov 20 	& (1409  - 3136) & Oguri et al., 2005 \\

HE 1104$-$1805  A, B  & 2.32  & 2008 Abr 07 	& (1310  - 2909) & Motta et al., 2012  \\

SDSS J1206+4332   A, B  & 1.789 & 2004 Jun 21 	& (1362  - 3048) & Oguri et al., 2005 \\

SDSS J1353+1138   A, B  & 1.629 & 2005 Apr 12 	& (1521  - 3385) & Inada et al., 2006 \\

WFI J2033$-$4723  B, C  & 1.66  & 2008 Apr 14 	& (1620  - 3625) & Motta, V., unpublished data  \\

HE 2149$-$2745  A, B & 2.033 & 2000 Nov 19 	& (1430  - 3174) & Motta, V., unpublished data  \\

\enddata

\end{deluxetable}

\clearpage

\begin{deluxetable}{c|c|c}

%\centering

\tablecaption{Wavelength regions\label{tbl-2}}

\startdata

\tableline

Feature & Wavelength intervals (\AA) & Description  \\

\tableline

Fe(1) 	&	$(\lambda2050\tablenotemark{1},\lambda2115)$ & \\	%	0047_ab_20020904.ods

Fe(2) 	&	$(\lambda2250,\lambda2320)\cup(\lambda2333,\lambda2445)\cup(\lambda2470,\lambda2625)$ &\\	

Continuum	&	$(\lambda2000,\lambda2020)\tablenotemark{2}$ & Bluewards of Fe(1) \\	%	Kuraszkiewicz et al. 2002

Continuum	&	$(\lambda2160,\lambda2180)\tablenotemark{3}$ & Redwards of Fe(1) \\	%	Kuraszkiewicz et al. 2002

Continuum 	&	$(\lambda2225,\lambda2250)\tablenotemark{3}$ & Bluewards of Fe(2)\\	%	Kuraszkiewicz et al. 2002

Continuum 	&	$(\lambda2640,\lambda2650)\tablenotemark{4}$ & Redwards of Fe(2)\\	%	2650 As suggested in Francis et al. 1991

%	0435_ab_20080112_MMT_vm.ods

Mg II$\lambda2798$ 	&($\lambda2776,\lambda2820$) & Line core\\ % 2798 +/-22, 

C III]$\lambda1909$ 	&($\lambda1893,\lambda1925$) & Line core\\ % 1909 +/- 16

\enddata

%\begin{footnote}

\tablenotetext{1}{Originally taken at $\lambda1942$ in Vestergaard et al. (2001).}

\tablenotetext{2}{Contaminated with the wing of the C III]$\lambda1909$ line (Kuraszkiewicz et al. 2002).}

\tablenotetext{3}{Pure continuum window according to Kuraszkiewicz et al. (2002).}

\tablenotetext{4}{As suggested in Francis et al. (1991). This continuum window is defined out of the Mg II$\lambda2798$ wings (Brotherton et al. 2001) and of the iron blend Fe(2).}

%\end{footnote}

\end{deluxetable}

\clearpage

\begin{deluxetable}{l|c|c|c}

\tablecaption{Differential microlensing using Mg II$\lambda2798$ as reference. \label{tbl-4}}

\startdata

\tableline

Object (pair) & Fe(1) region & Fe(2) region & $\lambda$2798 continuum  \\

\tableline

HE 0435$-$1223	(B-A)	&	$+0.17_{-0.04}^{+0.05}$ & $+0.37_{-0.04}^{+0.04}$ & $-0.26_{-0.03}^{+0.03}$ \\	%      0435_ab_20080112_MMT_vm.ods

HE 0435$-$1223	(D-B)	&	$+0.23_{-0.01}^{+0.01}$ & $+0.13_{-0.01}^{+0.01}$ & $+0.18_{-0.01}^{+0.01}$ \\	%      0435_bd_2004.10.12_VLT_vm.ods

SDSS J0806+2006	(B-A)	&	$+0.40_{-0.39}^{+0.63}$ & $+1.00_{-0.12}^{+0.13}$ & $+0.91_{-0.16}^{+0.19}$ \\	%      0806_ab_20050412.ods

SBS 0909+532	(B-A)	&	$-0.47_{-0.06}^{+0.07}$ & $-0.29_{-0.05}^{+0.05}$ & $+0.40_{-0.05}^{+0.05}$ \\	%      0909_azul_evencio.ods

SDSS J0924+0219	(B-A)	&	$+0.12_{-0.02}^{+0.02}$ & $-0.09_{-0.02}^{+0.02}$ & $-0.12_{-0.02}^{+0.02}$ \\	%      0924_ab_20050114.ods

FBQS J0951+2635	(B-A)	&	$+0.35_{-0.05}^{+0.05}$ & $+0.33_{-0.02}^{+0.02}$ & $+0.50_{-0.02}^{+0.02}$ \\	%      0951_ab_19970214.ods

QSO 0957+561	(B-A)	&	$-      	      $ & $+0.02_{-0.12}^{+0.14}$ & $+0.30_{-0.06}^{+0.06}$ \\	%      0957_ab_19990415_2000060x.ods

QSO 0957+561	(B-A)	&	$+0.57_{-0.04}^{+0.04}$ & $-0.02_{-0.04}^{+0.04}$ & $+0.66_{-0.04}^{+0.04}$ \\	%      0957_ab_20080113_mmt_vm.ods

SDSS J1001+5027	(B-A)	&	$+0.41_{-0.09}^{+0.10}$ & $+0.29_{-0.05}^{+0.05}$ & $-0.16_{-0.04}^{+0.04}$ \\	%      1001_ab_20031120.ods

HE 1104$-$1805	(B-A)	&	$-0.18_{-0.04}^{+0.04}$ & $-0.08_{-0.03}^{+0.03}$ & $-0.09_{-0.02}^{+0.02}$ \\	%      1104_ab_20080407_vlt_vm.ods

SDSS J1206+4332	(A-B)	&	$+0.50_{-0.09}^{+0.10}$ & $-0.37_{-0.15}^{+0.18}$ & $+0.39_{-0.10}^{+0.11}$ \\	%      1206_ab_20040621.ods

SDSS J1353+1138	(A-B)	&	$+1.04_{-0.03}^{+0.03}$ & $+0.94_{-0.02}^{+0.02}$ & $+0.00_{-0.02}^{+0.02}$ \\	%      1353_ab_20050412.ods

WFI J2033$-$4723	(B-C)	&	$-0.26_{-0.02}^{+0.02}$ & $-0.09_{-0.02}^{+0.02}$ & $-0.27_{-0.02}^{+0.02}$ \\	%      2033_bc_20080414_vm.ods

HE 2149$-$2745	(B-A)	&	$-0.03_{-0.03}^{+0.03}$ & $+0.13_{-0.02}^{+0.02}$ & $-0.02_{-0.02}^{+0.02}$ \\	%      2149_ab_xxxxxxxx_vm.ods

\enddata

\tablecomments{$\bigtriangleup m-\bigtriangleup m_{\rm Mg II\lambda2798}$, of the Fe(1) and Fe(2) blends after continuum subtraction, and of the Mg II$\lambda2798$ continuum. The Mg II$\lambda2798$ emission line flux (after continuum subtraction) is used as the no microlensing reference in all cases.}

\end{deluxetable}

\clearpage

\begin{deluxetable}{l|c|c|c}

\tablecaption{Differential microlensing using C III]$\lambda1909$ as reference. \label{tbl-5}}

\startdata

\tableline

Object (pair) & Fe(1) region & Fe(2) region & $\lambda$1909 continuum  \\

\tableline

HE 0047$-$1756	(B-A)	&	$-0.15_{-0.07}^{+0.08}$ & $-0.01_{-0.03}^{+0.03}$ & $+0.25_{-0.03}^{+0.03}$ \\%      0047_ab_20020904.ods

HE 0435$-$1223	(B-A)	&	$+0.32_{-0.03}^{+0.03}$ & $+0.52_{-0.02}^{+0.02}$ & $-0.21_{-0.02}^{+0.02}$ \\%      0435_ab_20080112_MMT_vm.ods

HE 0435$-$1223	(D-B)	&	$+0.18_{-0.02}^{+0.02}$ & $+0.08_{-0.02}^{+0.02}$ & $+0.23_{-0.02}^{+0.02}$ \\%      0435_bd_2004.10.12_VLT_vm.ods

SDSS J0806+2006	(B-A)	&	$+0.02_{-0.36}^{+0.55}$ & $+0.62_{-0.08}^{+0.08}$ & $+0.54_{-0.07}^{+0.08}$ \\%      0806_ab_20050412.ods

SBS 0909+532	(B-A)	&	$+0.54_{-0.10}^{+0.12}$ & $+0.72_{-0.09}^{+0.10}$ & $+0.66_{-0.11}^{+0.13}$ \\%      0909_azul_evencio.ods

SDSS J0924+0219	(B-A)	&	$+0.07_{-0.02}^{+0.02}$ & $-0.14_{-0.02}^{+0.02}$ & $-0.22_{-0.02}^{+0.02}$ \\%      0924_ab_20050114.ods

FBQS J0951+2635	(B-A)	&	$+0.72_{-0.07}^{+0.07}$ & $+0.70_{-0.04}^{+0.04}$ & $+0.92_{-0.04}^{+0.05}$ \\%      0951_ab_19970214.ods

QSO 0957+561	(B-A)	&	$+0.56_{-0.02}^{+0.02}$ & $-0.02_{-0.02}^{+0.02}$ & $+0.71_{-0.02}^{+0.02}$ \\%      0957_ab_20080113_mmt_vm.ods

SDSS J1001+5027	(B-A)	&	$+0.59_{-0.09}^{+0.10}$ & $+0.46_{-0.05}^{+0.05}$ & $-0.30_{-0.04}^{+0.04}$ \\%      1001_ab_20031120.ods

HE 1104$-$1805	(B-A)	&	$-0.16_{-0.08}^{+0.08}$ & $-0.06_{-0.06}^{+0.07}$ & $+0.01_{-0.06}^{+0.06}$ \\%      1104_ab_20080407_vlt_vm.ods

SDSS J1206+4332	(A-B)	&	$+0.43_{-0.08}^{+0.08}$ & $-0.44_{-0.14}^{+0.16}$ & $+0.50_{-0.08}^{+0.08}$ \\%      1206_ab_20040621.ods

SDSS J1353+1138	(A-B)	&	$+1.46_{-0.08}^{+0.08}$ & $+1.36_{-0.07}^{+0.08}$ & $+0.14_{-0.08}^{+0.08}$ \\%      1353_ab_20050412.ods

WFI J2033$-$4723	(B-C)	&	$-0.28_{-0.02}^{+0.02}$ & $-0.11_{-0.02}^{+0.02}$ & $-0.29_{-0.02}^{+0.02}$ \\%      2033_bc_20080414_vm.ods

HE 2149$-$2745	(B-A)	&	$+0.00_{-0.04}^{+0.04}$ & $+0.16_{-0.03}^{+0.03}$ & $-0.13_{-0.03}^{+0.03}$ \\%      2149_ab_xxxxxxxx_vm.ods

\enddata

\tablecomments{$\bigtriangleup m-\bigtriangleup m_{\rm C III]\lambda1909}$, of the Fe(1) and Fe(2) features after continuum subtraction, and of the C III]$\lambda1909$ continuum. The C III]$\lambda 1909$ emission line flux (after continuum subtraction) is used as the no microlensing reference in all cases.}

\end{deluxetable}

\end{document}